\documentclass[creativecommons,noderivs,noncommercial,copyright]{eptcs}
\providecommand{\event}{SYNT 2016}
\usepackage{color}
\usepackage{xspace}
\usepackage{ltl}
\usepackage{graphics}
\usepackage{sty/more-colors}
\usepackage{amssymb}
\usepackage{stmaryrd}
\usepackage{paralist}
\usepackage{amsthm}
\usepackage{sty/strat_shortcuts}
\newcommand{\fundinghint}{%
	This work has been partly supported by the German Research Foundation
	(DFG) as part of the Transregional Collaborative Research Center
	''Automatic Verification and Analysis of Complex Systems'' (SFB/TR 14 AVACS).}

\title{What You Really Need To Know About Your Neighbor\thanks{\fundinghint}}
\author{ Werner Damm
	\institute{CvO Universit\"at Oldenburg,\\ 26111 Oldenburg}	
	\email{werner.damm@offis.de}
	\and
	Bernd Finkbeiner
	\institute{Universit\"at des Saarlandes,\\Campus E1\,3, 66123 Saarbr\"ucken}
	\email{finkbeiner@cs.uni-saarland.de}
	\and
	Astrid Rakow
	\institute{CvO Universit\"at Oldenburg,\\ 26111 Oldenburg}	
	\email{a.rakow@uni-oldenburg.de}
	}

\def\titlerunning{What You Really Need To Know About Your Neighbor}
\def\authorrunning{W. Damm, B. Finkbeiner \& A. Rakow}

\begin{document}
\maketitle

\newcommand\admissible{\boxdotRight}
\newcommand\realizes{\boxRight}
\newcommand\stronglyadmissible{\DiamonddotRight_n}
\newcommand\outp{\mathit{out}}
\newcommand\inp{\mathit{inp}}

\newcommand\env{\mathit{env}}
\newcommand\comp{\mathit{comp}}
\newcommand\Venv{{V_{\mathit{env}}}}
\newcommand\dir{\mathit{dir}}
\newcommand\Alt{\mathit{Alt}}
\newcommand\child{\mathit{child}}

\def\Circle{\mathop{\hbox{\zmsy\char'062}}}
\def\Diamond{\mathop{\hbox{\zmsy\char'061}}}
\def\Box{\mathop{\hbox{\zmsy\char'060}}}

\newcommand{\dotcup}{\ensuremath{\mathaccent\cdot\cup}}

\begin{abstract}
	A fundamental question in system design is to decide how much of the design of one component must be known in order to successfully design another component of the system. We study this question in the setting of reactive synthesis, where one constructs a system implementation from a specification given in temporal logic. 
	In previous work, we have shown that the system can be constructed compositionally, one component at a time, if the specification admits a \emph{"dominant"} (as explained in Introduction) strategy for each component. 
%
	In this paper, we generalize the approach to settings where dominant strategies only exist under certain assumptions about the future behavior of the other components. We present an incremental synthesis method based on the automatic construction of such assumptions.
\end{abstract}
\section{Introduction}
We investigate the following fundamental research question in system design: Consider a system architecture $\archi$ composed of two processes $\proc_h$, $\proc_l$ of different priorities high and low, all with defined interfaces $\inp(\archi)$, $\outp(\archi)$, $\inp(\proc_h)$, $\outp(\proc_h)$, $\inp(\proc_l)$, $\outp(\proc_l)$ and objectives $\varphi_{\archi}$, $\varphi_{\proc_h}$, $\varphi_{\proc_l}$ expressed in linear time temporal logic using the interfaces of $\archi$, $\proc_h$, $\proc_l$. How much does $\proc_h$ need to know about $\proc_l$ so that 

\begin{enumerate}[(1)]
\item both jointly realize $\varphi_{\archi}$ whenever possible,
\item $\proc_h$ can realize its objective whenever possible,
\item $\proc_l$ only sacrifices achieving its objectives when this is the only way to achieve (1) and (2)?
\end{enumerate}
We provide a rigorous formulation of this research question, and give algorithms to compute the insight $\proc_h$ must have into $\proc_l$. We then show how this can be used to generate best-in-class strategies for $\proc_h$ and $\proc_l$, in the sense that if there is a strategy at all for $\proc_h$ which in the context of $\proc_l$ achieves both (1) and (2), then so will the best-in-class strategy for $\proc_h$, and if there is a strategy at all for $\proc_l$ which achieves (3), then so will the best-in-class strategy for $\proc_l$. The algorithm works for any number of processes with priorities henceforth represented by a partial order $<$.

Such system design questions arise naturally in many application areas, with $A$ being a system of systems, $\varphi_{A}$ representing overarching control objectives, and constituent systems $\proc_1, \ldots, p_n$ with their own local control objectives, where $<$ reflects the degree of criticality of the subsystem for the overall system. In such systems, the notion of neighborhood has its classical interpretation based on physical proximity. One example of such systems are smart grids, where subsystems represent different types of energy providers and consumers, and the highest priority subsystem is responsible for maintaining stability of the subgrid in spite of large fluctuations of inflow and outflow of energy. The running example we use comes from cooperative autonomous vehicle control, where the overall objectives relate to safety, avoiding congestion, fuel and $CO_2$ reduction etc, and local objectives relate to reaching destinations in given time frames, reducing fuel consumption, and executing certain maneuvers.  We focus here on simple objectives such as completing a maneuver with cooperation from neighboring cars.

As pointed out in \cite{damm+finkbeiner/2010/pay,df14}, for systems of systems controlling physical systems as in the two examples above, there will be no winning strategies, because rare physical events such as stemming from extreme weather conditions or physical system failures can occur, making it impossible to even achieve overarching safety objectives. No electronic stability system of a car would be able to still guarantee safety in unexpected icy road conditions in a dynamical situation with already stretched system limits. We introduced the notion of \emph{(remorse-free) dominant strategies} to compare strategies in such applications where a winning strategy cannot be found, making precise what has been called "best-in-class" strategies above. Such a strategy will -- for every sequence of environment actions -- be at least as good in achieving the system objectives as any other strategy. We can now make precise the addressed research question: how much insight does $\proc_h$ need into the plans of $\proc_l$, so that there is a dominant strategy for $\proc_h$ to achieve (1) and (2) above?

In general, local strategies must cater for needs of neighboring systems, so as to enable synthesis of  (remorse-free) dominant systems. Consider e.g. a two car system on the highway, with the \ego car wanting to reach the next exit, and having an \other,  neighboring car on the rightmost lane. There is no remorse-free dominant strategy of the \ego car to reach the exit, unless the \other car to its right cooperates: if the \ego car accelerates to overtake the car, the car on right might do the same; similarly, an attempt to reach the exit by slowing down might be blocked by the car to the right. Thus \ego needs insight into the future moves of the \other car. In fact, if the \other car would signal, that its next move will be to decelerate, then the \ego car can bet on accelerating so as to overtake the car and reach the exit. If, on the other hand, the \other car would promise to accelerate, then \ego could decelerate and thus change lane in order to reach the exit. The weakest assumption for \ego to have a dominant strategy will not only consider the next move of the \other car, but give the other car as much flexibility as possible, as long as a lane change will still be possible prior to reaching the exit. We formalize these as what we call \emph{assumption trees} of \ego, which branches at each point in time into the set of possible futures of settings of the speed control of the other car s.t. the \ego car has a dominant strategy in all such futures, and show that we can compute the most general assumption tree under which \ego has a dominant strategy. A strategy for the other car is then not only determining the speed control of other, but also picking only evolutions which are compatible with \ego's assumption tree. Rather than communicating the choice of such future moves through an extended interface, we provide \ego with a copy of the strategy of \other, so that \ego can determine the selected future by simulating the strategy of the \other car.

In general, such assumption trees can have unbounded depth. Consider the slight variant of the above example, where \ego only wants to change lanes (no exit targeted) and has as secondary objective to always keep its speed. The most general assumption tree of \ego will then, at each depth, have choices of futures where the \other car will always keep its speed, forcing \ego to either accelerate or decelerate to at least meet its higher priority objective of changing lane, and those futures where eventually \other either breaks or accelerates, in which case \ego can achieve both its objectives.

In general, we assume to have $n$ processes ordered by criticality $<$. Intuitively, if $\proc_h$ is more critical than $\proc_l$ ($\proc_h>\proc_l$), then we would want $\proc_l$ to adapt its behavior so as to be compliant to the assumption tree of $\proc_h$. For a totally ordered set of processes $\proc_1 > \proc_2 > \ldots > \proc_n$ we thus generate assumption trees inductively starting with $\proc_1$, and then propagate this along the total order. Assuming the overall environment of $\archi=\{\proc_1, \ldots, \proc_n\}$ meets the assumption tree of $\proc_n$, we can then synthesize for each $\proc_j$ a remorse-free dominant strategy meeting the canonical generalization of (1),(2),(3) above to $n$ processes. If processes are only partially ordered, we perform this algorithm for clusters of processes with same priority.

\paragraph{Related Work} We introduced the notion of remorse-free dominant strategies in \cite{damm+finkbeiner/2010/pay}, and provided methods for compositional synthesis of remorse-free dominant strategies in \cite{df14}. 
Compositional approaches have been studied in the setting of assume-guarantee synthesis \cite{AGS:2007}, where the synthesized strategies are guaranteed to be robust with respect to changes in other components, as long as the other components do not violate their own, local, specifications. In contrast, in this paper, we derive assumptions to ensure cooperation. 
The automatic synthesis of strategies for reactive systems goes back to the seminal works by Church, B\"uchi, Landweber, and Rabin in the 1960s \cite{Church/63/Logic,buechi1990,rabin1972automata}, and by Pnueli and Rosner in the 1980s \cite{Pnueli:1989}. For distributed systems, the synthesis problem is known to be undecidable in general \cite{Pnueli:1990}, and decidable under certain assumptions such as well-connected system architectures \cite{Kumar2006}.  
The problem that specifications usually must be strengthened with environment assumptions to become realizable has been recognized by several authors. 
The construction by Chatterjee et al \cite{Chatterjee2008} directly operates on the game graph. 
A safety assumption is computed by removing a minimal set of environment edges from the graph. 
A liveness assumption is then computed by putting fairness assumptions on the remaining edges. 
Liu et al \cite{Liu2011} mine counterexamples to realizability to find new assumptions. 
A key difference between the assumptions generated in these approaches and the assumptions of this paper is that we compute assumptions for the existence of dominant strategies, not for the existence of winning strategies. As a result, our assumptions are weaker. 
In particular, it is in general not necessary to restrict the behavioral traces of the other components, only their branching structure.

The notion of admissibile strategy in \cite{Berwanger2007,Faella2007,Faella2009,brenguier2015}  is closely related to our notion of dominant strategy. 
In \cite{Berwanger2007} Berwanger presents a method based on iterated admissibility and gives existence results for general multi-player games of infinite duration. Faella presents in \cite{Faella2009} an efficient way to compute admissible strategies by computing the conventional winning strategy, which is applied from winning states, and a cooperatively winning strategy for the remaining states.
Brenguier et al \cite{brenguier2015} introduce an approach called assume-admissible synthesis.
In assume-admissible synthesis it is assumed a-priori that the other components also apply admissible strategies, while we make no such assumption.

To use formal methods and in particular formal synthesis methods for coordinated vehicle maneuvers has been proposed in among others \cite{Frese2010,ies_2010_frese_tr,Wongpi2013,MSC:8951470,DPR-IJFCS07,Dzetkilc2011,Wongpi2012,Mickelin2014}. \cite{Frese2010} searches for strategies controlling all vehicles, and employs  heuristic methods from artificial intelligence such as tree-search to determine strategies for coordinated vehicle movements. An excellent survey for alternative methods for controlling all vehicles to perform collision free driving tasks is given in \cite{ies_2010_frese_tr}. Both methods share the restriction of the analysis to a small number of vehicles, in contrast to our approach, which is based on safe abstractions guaranteeing collision freedom in achieving the driver's objectives taking into account the complete traffic situation.  In \cite{Coogan2014}, the authors show that LTL formulas are expressive enough to express typical traffic-flow optimization objectives, and use formal synthesis methods to automatically generate control commands for access control on highway systems. In \cite{Wongpi2013}, the authors describe a methodological approach for decomposing the synthesis problem, which, however, has already been described in a previous journal publication by the authors \cite{MSC:8951470}. It additionally presents a tool for constructing the finite abstraction; our previous work in \cite{DPR-IJFCS07,Dzetkilc2011} goes beyond the capabilities of this tool in also allowing for non-linear dynamics. In \cite{Wongpi2012}, the authors provide an improved two-level approach for control synthesis that however suffers from a flaw in constructing the abstraction relation. In \cite{Mickelin2014} the authors provide robust finite abstractions with bounded estimation errors for reducing the synthesis of winning strategies for LTL objectives to finite state synthesis and demonstrate the approach for an aerospace control application; however, this approach does not cover cooperative maneuvers. We finally mention that in the aerospace domain, high-level objectives such as maximizing throughput and energy efficiency while maintaining safety have led to new concepts for air-traffic control, such as free flight. There is a significant body of work on the verification of the safety of such control strategies as well as on design rules that ensure safety \cite{Tomlin+others/98/Advanced,Henzinger1998,Lygeros1997,Richards2002,ADEFH+01,DammDHJPPSWW2007}. 

\paragraph{Outline} Our paper is structured as follows. We introduce preliminaries on strategy synthesis and tree automata in Section \ref{sect:synthesis} and \ref{sect:preliminaries}. Section \ref{sect:assumptions} shows how to compute assumptions for cooperation. Section \ref{sect:inc_strategies} provides the incremental distributed synthesis procedure.
As a running example to illustrate our notions, we use the \emph{side by side} example, where a car A starts side by side with car B and A's  goal is to eventually be before or after B.
We conclude in Sect.~\ref{sect:conclusion}.
\section{Synthesis of Distributed Systems}\label{sect:synthesis}
We consider complex multi-component systems at an early design stage, where the system architecture and the design objectives are known, but the components have not been implemented yet.
We are interested in synthesizing an implementation for a given system architecture $\archi$ and a specification $\spec$.  A solution to the synthesis problem is a set of finite-state strategies $\{s_\proc \mid \proc \in \procSet\}$, one for each process in the architecture, such that the joint behavior satisfies $\spec$.

\paragraph{Architectures} An \emph{architecture} $\archi$ is a tuple $(\procSet,V,\inp,\outp)$, where $\procSet$ is a set of system processes,
$V$ is a set of (Boolean) variables, and $\inp, \outp: \procSet \rightarrow 2^V$ are two functions that map each process to a set of input and output variables, respectively. For each process $\proc$, the inputs and outputs are disjoint, $\inp(\proc) \cap \outp(\proc) = \emptyset$, and for two different processes $\proc \neq q$, the output variables are disjoint: $\outp(\proc) \cap \outp(q) = \emptyset$. We denote the set of visible variables of process $\proc$ with $V(\proc) = \inp(\proc) \cup \outp(\proc)$.
Variables $V_I = V \smallsetminus \bigcup_{\proc\in \procSet} \outp(\proc)$ that are not the output of any system process are called \emph{external inputs}. We assume that the external inputs are available to every process, i.e., $V_I \subseteq \inp(\proc)$ for every $\proc \in \procSet$. 

For two architectures $\archi_1 = (\procSet_1,V,\inp_1,\outp_1)$ and $\archi_2 = (\procSet_2,V,\inp_2,\outp_2)$ with the same variables, but disjoint sets of processes, $\procSet_1 \cap \procSet_2 = \emptyset$, we define the parallel composition as the architecture ${\archi_1 \parallel \archi_2} = (\procSet_1 \cup \procSet_2, V, \proc \mapsto \mbox{if } \proc \in \procSet_1 \mbox{ then } \inp_1(\proc) \mbox{ else } \inp_2(\proc),  \proc \mapsto \mbox{if } \proc \in \procSet_1 \mbox{ then } \outp_1(\proc) \mbox{ else } \outp_2(\proc))$.

\begin{exampleblock}
We use as running example the scenario from the introduction, where a car is side by side to another car and has to reach an exit. To be usable as running example, we strip this down to the bare essentials.
Our formal model will consider this two car system as an architecture consisting of two "processes",  which we call \ego and \other. 
We restrict ourselves to finite state systems, and abstract the physical notion of positions of the two cars on adjacent lanes of a highway into the states \emph{\sbs}, and \emph{\nsbs}, which are interpreted from the perspective of the \ego car. 
Each car can control its own dynamics by accelerating, keeping the current speed, or decelerating. These actuators are captured in our formal model as output variables. 
E.g. for the process \ego its output variables $\outp(\ego)$ are $\accel_e$, $\keep_e$, $\decel_e$ with the obvious meaning. 
We consider fully informed processes, hence \ego has $\outp(\other)$ as input variable. 

We capture the effect of the setting of output variables and input variables of a process on the system state in what we call \emph{world models}. 
Figure~\ref{fig:world_model_side_by_side} shows the world model of the ego car.
Initially, both cars are assumed to be side by side. State transitions are labeled by pairs of the chosen output of the process and the observed input of the process. 
E.g. if \ego chooses to keep its speed, and the \other car keeps its speed, too, then the system remains in state \sbs. If, on the other hand, \ego chooses to accelerate, while \other  keeps its velocity, the system transitions to state \nsbs. 
 Multiple labels are just shorthand for multiple transitions with unique labels. Note that the world model of \other is identical to that of \ego except for exchanging inputs and outputs. From a control-theoretic perspective, the relative position of cars is a plant variable, which is influenced by the actuators of both cars. 
\begin{figure}
\centering
\includegraphics[width=0.8\textwidth]{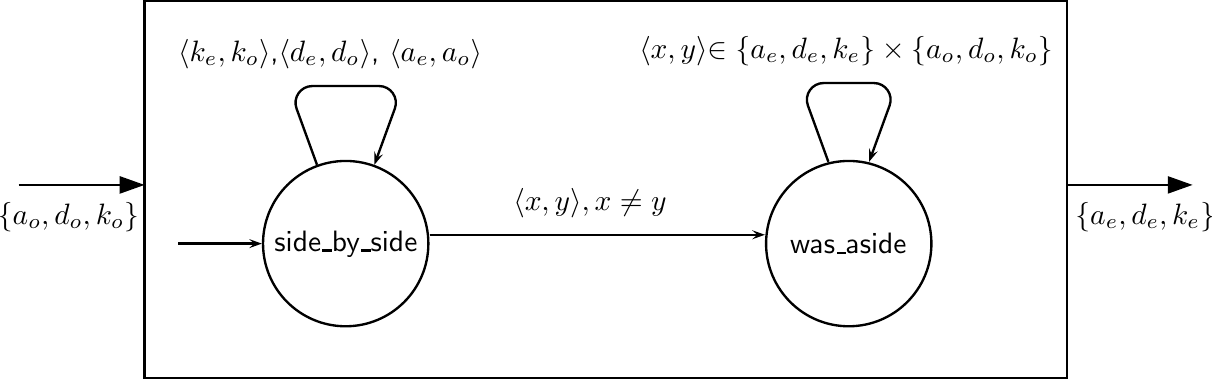}

\caption{World model}\label{fig:world_model_side_by_side}
\end{figure}
\end{exampleblock}

We now give the formal definition of the world model of a process $\proc$. 

Let $\sysStates$ be a set of (system-) states. A world model $\world$ for process $\proc$ with input variables $\inp(\proc)$ and output variables $\outp(\proc)$ is an input deterministic transition system over $\sysStates$ with designated initial state and transitions labeled by pairs $\langle o,i\rangle$ where $i\in \inp(\proc)$ and $o \in \outp(\proc)$. Formally
$\world(\proc) = (\sysStates, E, L, s_0 )$
with $E \subseteq  \sysStates \times \sysStates$, $L: E \rightarrow \outp(p) \times \inp(p)$, $s_0 \in \sysStates$ s.t. 
$(s,s_1)\in E \land (s,s_2)\in E \land L(s,s_1)=L(s,s_2) \Rightarrow s_1=s_2$
\paragraph{Implementations}  An \emph{implementation} of an architecture consists of
strategies $\stratSet = \{s_{\proc} \mid \proc \in \procSet\}$ for the system processes.
A system process $\proc \in \procSet$ is implemented by a \emph{strategy},
i.e., a function $s_{\proc}:(2^{\inp(\proc)})^{*} \rightarrow 2^{\outp(\proc)}$ that maps histories of inputs to outputs. 
A strategy is \emph{finite-state} if it can be represented by a finite-state \emph{transducer}  $(Q,q_{0},\delta:Q \times 2^{\inp(\proc)} \rightarrow Q, \gamma: Q \rightarrow 2^{\outp(\proc)})$, with a finite set of states $Q$, an initial state $q_{0}$, a transition function $\delta$ and an output function $\gamma$.

The parallel composition $s_p || s_q$ of the strategies of two processes $\proc,q \in P$ is a function $s_{p||q}: (2^{I})^* \rightarrow 2^{O}$  that maps histories of the remaining inputs $I = (\inp(p) \cup \inp(q)) \setminus (\outp(p) \cup \outp(q))$ to the union $O = \outp(p) \cup \outp(q)$ of the outputs:
$s_{p||q}(\sigma) = s_p(\alpha_p(\sigma)) \cup s_q(\alpha_q(\sigma))$, where
$\alpha_p : (2^I)^* \rightarrow (2^{\inp(p)})^*$ fills in
for process $p$ what $q$ contributes to the $p$'s input, that is
$\alpha_p(\epsilon) = \epsilon$ and
$\alpha_p(\upsilon_0\upsilon_1\ldots\upsilon_k) = ((\upsilon_0 \cup s_q(\epsilon)) \cap \inp(p))  ((\upsilon_1 \cup s_q(\alpha_q(\upsilon_0))) \cap \inp(p)) \ldots
((\upsilon_k \cup s_q(\alpha_q(\upsilon_1 \upsilon_2 \ldots \upsilon_{k-1}))) \cap \inp(p))$,
and, analogously, $\alpha_q(\epsilon) = \epsilon$ and
$\alpha_q(\upsilon_0\upsilon_1\ldots\upsilon_k) = ((\upsilon_0 \cup s_p(\epsilon)) \cap \inp(q))  ((\upsilon_1 \cup s_p(\alpha_p(\upsilon_0))) \cap \inp(q)) \ldots
((\upsilon_k \cup s_p(\alpha_p(\upsilon_1 \upsilon_2 \ldots \upsilon_{k-1}))) \cap \inp(q))$.

A \emph{computation} is an infinite sequence of variable valuations. 
For a  sequence $\gamma=\upsilon_1\upsilon_2\ldots \in (2^{V \smallsetminus \outp(p)})^\omega$ of valuations of the variables outside the control of a process~$\proc$, the computation resulting from $s$ is denoted by $\mathit{comp}(s, \gamma) = (s(\epsilon)\cup \upsilon_1)\, (s(\upsilon_1 \cap {\inp(p)})\cup \upsilon_2)\, (s(\upsilon_1\cap{\inp(p)}\upsilon_2\cap{\inp(p)})\cup \upsilon_3)\ldots$.


\paragraph{Specification} We use linear-time temporal logic (LTL) to specify properties. In the following, we will denote the \texttt{next-time} operator with $\varbigcirc$, \texttt{globally} with $\LTLsquare$ and \texttt{eventually} with $\LTLdiamond$. For a computation $\sigma$ and an $\omega$-regular language $\varphi$, we write  $\sigma \models \varphi$ if $\sigma$ satisfies $\varphi$.
Design objectives are often given as a \emph{prioritized} conjunction of LTL formulas $\varphi = \varphi_1 \wedge \varphi_2 \wedge \ldots \wedge \varphi_n$, where $\varphi_1$ is the most important objective and $\varphi_n$ is the least important objective. For a priority $k$, with $1\leq k \leq n$, we consider the partial conjunction $\varphi^{k} = \bigwedge_{i=1\ldots k} \varphi_i$ and say that 
$\sigma$ satisfies $\varphi$ \emph{up to priority $k$} if $\sigma \models \varphi^{k}$.

\begin{exampleblock} 
	In the side by side example, the control objective of the \ego car is to reach a state where it can change lane, i.e., to eventually drive the plant into a state different from \sbs. From the perspective of \ego, the actuator settings of the \other car are uncontrollable disturbances. Suppose that the \ego car has the secondary objective to reduce fuel consumption, which very abstractedly can be captured by avoiding accelerating and decelerating altogether, in other words to always output $\keep_e$.  Suppose finally that \other is owned by a driver who may avoid getting tired by changing speed every now and then. We capture these overall control objectives as list of prioritized LTL formulas:
\begin{compactenum}
\item\label{specOne} $\LTLdiamond$ $\neg$ side$\_$by$\_$side 
\item $\LTLsquare$ $k_e$
\item $\LTLsquare \keep_o \lor \LTLsquare \LTLdiamond a_o \land \LTLsquare \LTLdiamond d_o$.
\end{compactenum}

The synthesis problem, we then are solving, is to find strategies for \ego and \other which ultimately allow \ego to change lane and avoid violating the lower priority objectives. 
\end{exampleblock}

A strategy $s_\proc: (2^I)^* \rightarrow 2^{O}$ is \emph{winning} for a property $\varphi$, denoted by $s_{\proc}\models \varphi$, iff, for every sequence $\gamma=\upsilon_1\upsilon_2\ldots \in (2^{V \smallsetminus O})^\omega$ of valuations of the variables outside the control of $\proc$, the computation $\mathit{comp}(s_{\proc}, \gamma)$ resulting from $s_{\proc}$ satisfies~$\varphi$.
We generalize the notion of winning from strategies to implementations (and, analogously, the notions of dominance and bounded dominance later in the paper), by defining that
an implementation $\stratSet$ is winning for $\varphi$  iff the parallel composition of the strategies in $\stratSet$ is winning (for their combined sets of inputs and outputs).
\paragraph{Synthesis}
A property $\varphi$ is
\emph{realizable} in an architecture $\archi$ iff there exists an implementation 
that is winning for $\varphi$. 

\section{Preliminaries: Automata over Infinite Words and Trees}
\label{sect:preliminaries}

We assume familiarity with automata over infinite words and trees. In
the following, we only give a quick summary of the standard
terminology, the reader is referred to \cite{E01} for a full
exposition.

A \emph{full tree} is given as the set $\Upsilon^*$ of all finite
words over a given set of directions $\Upsilon$.  For given finite
sets $\Sigma$ and $\Upsilon$, a $\Sigma$-\emph{labeled}
$\Upsilon$-\emph{tree} is a pair $\langle \Upsilon^*,l \rangle$ with a
labeling function $l: \Upsilon^* \rightarrow \Sigma$ that maps every
node of $\Upsilon^*$ to a letter of~$\Sigma$.

An \emph{alternating tree automaton} $\mathcal{A} = (\Sigma,\Upsilon,Q,q_0,\delta,\alpha)$
runs on full $\Sigma$-labeled $\Upsilon$-trees.
$Q$ is a finite set of states, $q_0 \in Q$ a designated initial state,
$\delta$ a transition function $\delta: Q \times \Sigma \rightarrow
\mathbb B^+(Q \times \Upsilon)$, where $\mathbb B^+(Q \times
\Upsilon)$ denotes the positive Boolean combinations of $Q \times
\Upsilon$, and $\alpha$ is an acceptance condition.  Intuitively,
disjunctions in the transition function represent nondeterministic
choice; conjunctions start an additional branch in the run tree of the
automaton, corresponding to an additional check that must be passed by the input tree.
A \emph{run tree} on a given $\Sigma$-labeled $\Upsilon$-tree $\langle
\Upsilon^*,l \rangle$ is a $Q \times \Upsilon^*$-labeled tree where
the root is labeled with $(q_0,l(\varepsilon))$ and where for a node
$n$ with a label $(q,x)$ and a set of children $\child(n)$, the labels
of these children have the following properties:
\begin{itemize}
  \item for all $m \in \child(n):$ the label of $m$ is $(q_m,x \cdot
  \upsilon_m)$, $q_m \in Q, \upsilon_m \in \Upsilon$ such that $(q_m,
  \upsilon_m)$ is an atom of $\delta(q,l(x))$, and
  \item the set of atoms defined by the children of $n$ satisfies $\delta(q,l(x))$.
\end{itemize}
A run tree is \emph{accepting} if all its paths fulfill the acceptance condition.
A \emph{parity condition} is a function $\alpha$ from $Q$ to a finite set of
colors $C \subset \mathbb{N}$. A path is accepted if the highest color
appearing infinitely often is even. The \emph{safety condition} is the special case
of the parity condition where all states are colored with $0$. The \emph{B\"uchi condition}
is the special case of the parity condition where all states are colored with either $1$ or $2$,
the \emph{co-B\"uchi condition} is the special case of the parity condition where all states are colored with either $0$ or $1$.
For B\"uchi and co-B\"uchi automata we usually state the coloring function in terms of a set $F$ of states.
For the B\"uchi condition, $F$ contains all states with color $2$ and is called the set of \emph{accepting} states. For the co-B\"uchi condition, $F$ contains all states with color $1$ and is called the set of \emph{rejecting} states. The B\"uchi condition is satisfied if some accepting state occurs infinitely often, the co-B\"uchi condition is satisfied if all rejecting states only occur finitely often.
A $\Sigma$-labeled $\Upsilon$-tree is \emph{accepted} if it has an accepting run tree.
The set of trees accepted by an alternating automaton $\mathcal{A}$ is called
its \emph{language} $\mathcal{L}(\mathcal{A})$. An automaton is empty iff its
language is empty. In addition to full trees, we also consider \emph{partial trees}, which are given as prefix-closed subsets of  $\Upsilon^*$. As partial trees can easily be embedded in full trees (for example, using a labeling $\Upsilon^* \mapsto \mathbb B$ that indicates whether a node is present in the partial tree), tree automata can also be used to represent sets of partial trees.

A \emph{nondeterministic} automaton is an alternating automaton where the
image of $\delta$ consists only of such formulas that, when rewritten in disjunctive
normal form, contain at most one element of $Q \times \{\upsilon\}$ for every direction $\upsilon$ in every
disjunct. A \emph{universal} automaton is an  alternating automaton where the image of $\delta$ contains no disjunctions.
A \emph{deterministic} automaton is an alternating automaton that is both universal and nondeterministic, i.e., the image of $\delta$ has no disjunctions and contains at most one element of $Q \times \{\upsilon\}$ for every direction $\upsilon$.

A \emph{word automaton} is the special case of a tree automaton where the
set $\Upsilon$ of directions is singleton. For word automata, we omit the direction
in the transition function.

\section{Dominant Strategies}
\label{sec:dominating}

In previous work, we introduced the notion of
\emph{remorse-free dominance}~\cite{damm+finkbeiner/2010/pay} in order to deal with situations where it is impossible to achieve the specified objective.
Dominance is a weaker version of winning. 
A strategy  $t: (2^I)^* \rightarrow 2^{O}$  \emph{is dominated by} a strategy  $s: (2^I)^* \rightarrow 2^{O}$, 
denoted by $t \preceq s$, iff, for every sequence $\gamma \in (2^{V \smallsetminus O})^\omega$ for which the computation $\mathit{comp}(t, \gamma)$ resulting from $t$ satisfies the objective specification up to priority
$m$, the computation $\mathit{comp}(s, \gamma)$ resulting from $s$ satisfies the objective specification up to priority $n$ and $m\leq n$.
A strategy $s$ is  \emph{dominant} iff, for all strategies  $t$, $t \preceq s$. Analogously to the definition of winning implementations, we say that an implementation $\stratSet$ is dominant iff the parallel composition of the strategies in $\stratSet$ is dominant.

Finally, we say that a property $\varphi$ is \emph{admissible} in an architecture $\archi$ iff there is a dominant implementation.
Informally, a specification is admissible if the question whether it can be satisfied does not depend on variables that are not visible to the process or on \emph{future inputs}. For example, the specification $\varphi = (\LTLcircle a) \leftrightarrow b$, where $a$ is an input variable and $b$ is an output variable is not admissible, because in order to know whether it is best to set $b$ in the first step, one needs to know the value of $a$ in the second step. No matter whether the strategy sets $b$ or not, there is an input sequence that causes \emph{remorse}, because $\varphi$ is violated for the chosen strategy while it would have been satisfied for the same sequence of inputs if the other strategy had been chosen.

\begin{theorem}\cite{df14}
For a process $\proc$, one can construct a parity tree automaton such that the trees accepted by the automaton define exactly the dominant strategies of $\proc$. 
\end{theorem}
\begin{exampleblock}
In the side by side example, the specification is not admissible:
Consider $\stratKEEP^{\omega}$, a strategy where \ego always keeps its speed, i.e., applies $\keep_e^{\omega}$. Following $\stratKEEP^{\omega}$, \ego will achieve objectives 1 and 2 on p.~\pageref{specOne}, if \other accelerates or decelerates at least once. But in case the \other car will always keep its speed, following $\stratKEEP^{\omega}$ \ego will not even achieve its most important objective, $\LTLdiamond \neg\sbs$. In this case though, following strategy $\stratACC$, which is to accelerate at the first step and then keep speed, \ego will achieve objective 1.  So, neither \stratACC nor $\stratKEEP^{\omega}$ is dominant.
Ego will feel remorse, if it follows strategy \stratACC but \other eventually chooses $\accel_o$ or $\decel_o$, and ego will also feel remorse, if it follows ${\stratKEEP^{\omega}}$ while \other also always keeps its speed. In order to know whether to choose ${\stratKEEP^{\omega}}$ or \stratACC, \ego needs to know whether \other will eventually change its speed or always keep speed.
\end{exampleblock}
%
\section{Computing cooperation assumptions}\label{sect:assumptions}
%
A component does not necessarily have a dominant strategy, even if the composite system has a dominant strategy. Suppose that, in the example specification $\varphi = (\LTLcircle a) \leftrightarrow b$ discussed above, the architecture consists of two processes $\proc$ and $q$, with outputs $b$ and $a$, respectively. Clearly, the system has a dominant, and even winning strategy, e.g., always set $a$ and $b$ to $\mathit{false}$, while $\proc$, as discussed above, does not have a dominant strategy. An interesting observation is that $\proc$ does have a dominant strategy if it is guaranteed that $q$ will set $a$ to $\mathit{false}$ in the next step, and that it also has a dominant strategy if it is guaranteed that $q$ will set $a$ to $\mathit{true}$ (in which case, $\proc$ would set $b$ to $\emph{true}$ as well). 

We formalize such assumptions on the environment behavior that guarantee the existence of a dominant strategy as \emph{assumption trees}. An \emph{assumption tree} for process $\proc$ is a subtree of the full $\mathbb N \cup 2^{(V \setminus \outp(p))}$-tree where, on every path, the direction alternates between $\mathbb N$ and $2^{V \setminus \outp(p)}$. Intuitively, each number corresponds to a promise of the environment to a particular restriction on its future behavior. The values of $2^{V \setminus \outp(p)}$ are then the execution of this promise. The future behavior following different numbers can differ in the inputs provided by the environment at specific points in time in the future and also in the number of branchings with natural numbers that occur in the future. 

\begin{exampleblock}
	In our side by side example, a very restrictive assumption is \emph{"\other always chooses $\keep_o$"}. This assumption can be envisioned as a tree that degenerates to a simple linear graph as in Fig.~\ref{fig:sbs_assm_ego}\,a). At the root the single edge announces that only one future is assumed initially. From there the single edge labeled $\keep_o$ encodes that the only allowed first step of this future is $\keep_o$. Likewise, the single edge at node $1\,\keep_o$ announces that only one future development is distinguished at that point. The single edge at $1\,\keep_o\,1$ encodes that the next step has to be $\keep_o$ again, and so on.
\begin{figure}
\centering
\begin{minipage}{\textwidth}
\begin{minipage}{0.33\textwidth}
	\includegraphics[scale=1.1]{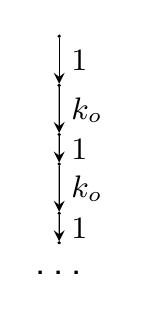}
\end{minipage}
\hfill
\begin{minipage}{0.63\textwidth}
	\includegraphics[scale=0.9]{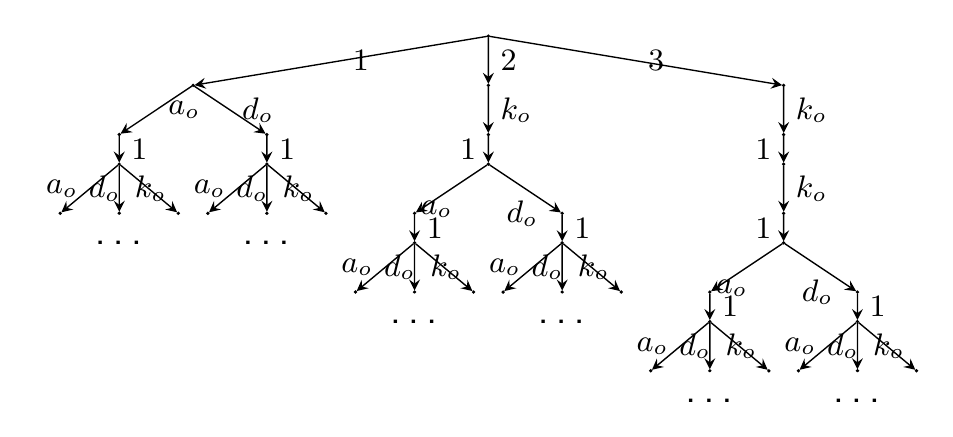}
\end{minipage}
\end{minipage}
\begin{minipage}{\textwidth}
\begin{minipage}{0.35\textwidth}
	{a) \ego assumes, \other does $\stratKEEP^\omega$}
\end{minipage}
\hfill
\begin{minipage}{0.6\textwidth}
		{b) \ego assumes, \other changes speed with the first three steps
	}
\end{minipage}
\end{minipage}
\caption{Two assumption trees of \ego}\label{fig:sbs_assm_ego}
\end{figure}
With respect to this assumption, \ego cannot achieve ${\LTLdiamond \neg \sbs} \land {\LTLsquare \keep_e}$, but \ego can achieve objective 1 without remorse following any strategy that chooses at least once $\accel_e$ or $\decel_e$. Any such strategy is dominant wrt. the assumption tree in Fig.~\ref{fig:sbs_assm_ego}\,a). 

Ego may also assume that \emph{"\other eventually chooses $\neg\keep_o$"}, where $\neg\keep_o$ abbreviates $\accel_o\lor\decel_o$. The strategy $\stratKEEP^{\omega}$ is the only dominant strategy with respect to this assumption and any stronger assumption -- including the one depicted in Fig.~\ref{fig:sbs_assm_ego}\,b). 
The tree of  Fig.~\ref{fig:sbs_assm_ego}\,b) roughly describes \ego's assumption that \other changes its speed within the first three steps.
At the tree's root, three assumptions on \other's behavior are distinguished. 
The first, at node $1$, describes that \other chooses initially $\neg \keep_o$. 
The subtrees at $1\,\accel_o$ and $1\,\decel_o$, respectively, both describe the assumption that all future development is possible:
The outgoing single edge at node $1\,\accel_o$ announces that then only a single future is distinguished.
The three outgoing edges at $1\,\accel_o 1$ specify that this future assumption allows to apply any acceleration, i.e., $\accel_o$, $\decel_o$ or $\keep_o$, as second step.
From there again, but not depicted, a single edge leads to a node with three outgoing edges labeled $\accel_o$, $\decel_o$ and $\keep_o$ denoting that any choice for \other's acceleration is accepted for this step as well; and so on.
The subtree at node $2$ in Fig.~\ref{fig:sbs_assm_ego}\,b) describes that \other initially chooses $\keep_o$, then $\neg \keep_o$ and thereafter the future is unconstrained. 
The subtree at node $3$ in Fig.~\ref{fig:sbs_assm_ego}\,b) describes that \other is assumed to choose $\keep_o\,\keep_o\,$ followed by $\neg\keep_o$ and then the future is no further constrained. 

For an example of an assumption tree with infinite branching at the root, consider \ego's assumption that \other announces in subtree $i$ to do $\neg \keep_o$ at step $i\in\mathbb N$.
\end{exampleblock}

To ensure that a given assumption tree guarantees the existence of a dominant strategy, we annotate each node that is reached by a natural number with an output value from $2^{\outp(p)}$. One such strategy annotation $t$ is \emph{dominated} by another strategy annotation $s$ of the same assumption tree, denoted by $t \preceq s$, iff for every alternating sequence of natural numbers and environment outputs from  $2^{\outp(p)}$ that is present in the tree, if the computation resulting from $t$ satisfies the objective specification up to priority
$m$, then the computation resulting from $s$ satisfies the objective specification up to priority $n$ and $m\leq n$. A strategy annotation $s$ is \emph{dominant} iff, for all strategies  $t$, $t \preceq s$. We say that an assumption tree is \emph{dominant} (and it \emph{guarantees the existence of a dominant strategy}) iff it has a dominant strategy annotation.

\begin{exampleblock} In Fig.~\ref{fig:sbs_assm_ego}\,b) where the assumption tree basically expresses that \other guarantees to choose $\neg \keep_o$ initially or within the next two steps, we can annotate the assumption tree with strategy $\stratKEEP^{\omega}$ to get a dominant strategy annotation. 

An example of a strategy annotated assumption tree is shown in Fig.~\ref{fig:sbs_strat+assm_keep}. There we basically combined the assumptions \emph{"\other chooses always $\keep_o$"} and \emph{"\other guarantees to choose $\neg\keep_o$ initially or within its next two steps"} as subtrees 1 and 2, respectively. The strategy annotation results from applying the strategy that alternates $\accel_e$ and $\decel_e$, $(a_ed_e)^\omega$ on subtree 1, and the $\stratKEEP^\omega$ strategy on subtree 2.  
\begin{figure}
\centering
\includegraphics[scale=0.8]{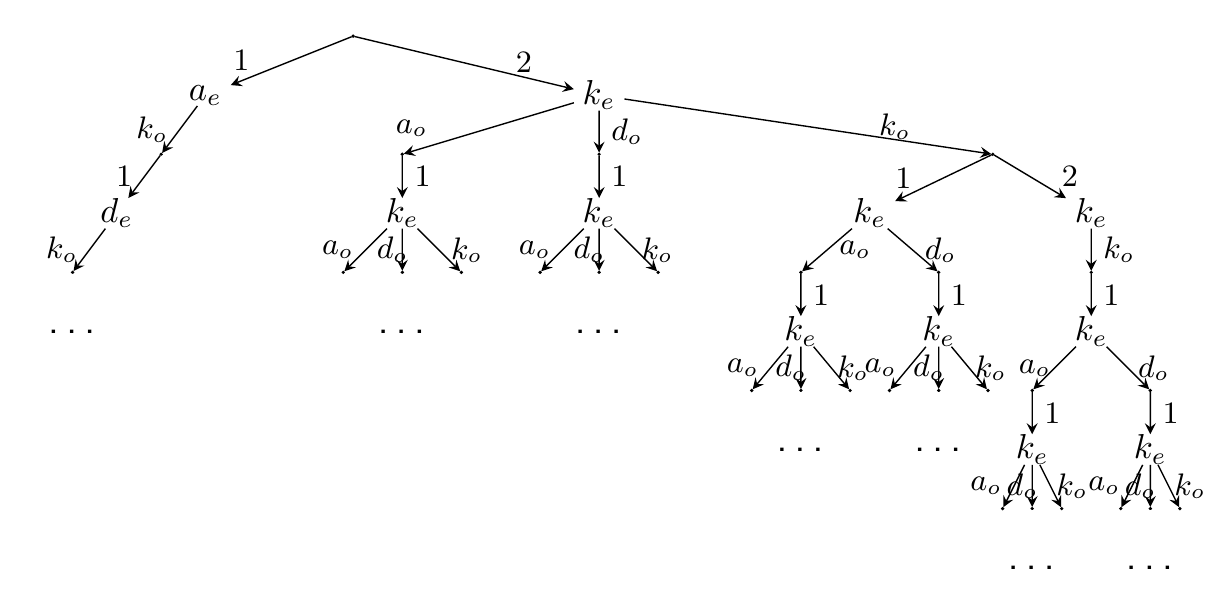}
\caption{\Sbs example: A strategy annotated assumption tree of ego.}\label{fig:sbs_strat+assm_keep}
\end{figure}
\end{exampleblock}

We now describe the computation of a tree automaton that recognizes the set of all assumption trees for a given process with a dominant annotation.
A small technical difficulty is that assumption trees have an infinite branching degree. In order to represent the set of assumption trees as a tree automaton, we use the standard unary representation of $\mathbb N^*$ in $\mathbb B^*$, where a sequence $abc\ldots \in \mathbb N^*$ is encoded as $0^a10^b10^c1\ldots$. The construction then proceeds in the following steps.
(1) We build a word automaton $\mathcal A$ that checks that no alternative would do better on the present path.
(2) We build a tree automaton $\mathcal B$ that reads in a strategy annotated assumption tree and uses $\mathcal A$ to verify that this annotation is dominant.

\begin{theorem}

\label{theo:assumptiontree}
For a single process $\proc$, one can construct a parity tree automaton such that the trees accepted by the automaton are the dominant strategy annotated assumption trees of $\proc$.
\end{theorem}
%
\section{Incremental synthesis of cooperation strategies}\label{sect:inc_strategies}
%
We now use the construction of the environment assumptions from Section~\ref{sect:assumptions} to incrementally synthesize a distributed system by propagating the assumptions.
We assume that the processes $P=\{p_1, p_2, \ldots, p_n\}$ are ordered
$\proc_1 > p_2 > \ldots p_n$ according to criticality.
We begin by computing the set $A_1$ of strategy annotated assumption trees for process $\proc_1$. Next, we compute the set $A_2$ of strategy annotated assumption trees for process $\proc_2$ that additionally satisfy $A_1$. We continue up to process $\proc_n$.

\begin{theorem}
For an architecture with a set  $P=\{\proc_1, \proc_2, \ldots, \proc_n\}$ of processes,  ordered
$\proc_1 > \proc_2 > \ldots \proc_n$ according to criticality,
one can construct, for each process $\proc_i, i=1..n$, a parity tree automaton, such that the trees accepted by the automaton are assumption trees that guarantee the existence of dominant strategies for all $\proc_j$ with $j\leq i$.
\end{theorem}

The set of assumption trees constructed in this way represent the remaining assumptions on the external inputs. Since we cannot constrain the external environment, we find the final distributed implementation by selecting some assumption tree that allows for all possible future environment behaviors.

\begin{exampleblock}
\begin{figure}
\centering
\includegraphics[scale=0.8]{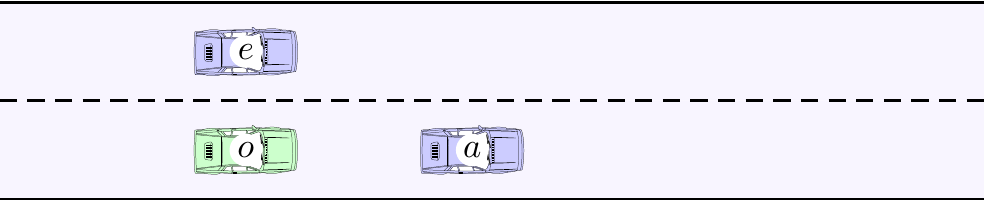}
\caption{Initial scenario for the incremental synthesis: Three cars on a freeway.}\label{fig:init}
\end{figure}
Let us consider three cars as in Fig.~\ref{fig:init}. Ego and \other are initially side by side and the car \ahead starts in front of \other. The synthesis problem is to find a strategy for \ego, \other and \ahead that satisfies the prioritized specification given by the following objectives listed in order of decreasing priority.
\begin{enumerate}
	\item ${\LTLdiamond \neg \sbs_{e}} \land {\LTLsquare (\bend \Rightarrow \LTLcircle \LTLcircle \neg \accel_a)}$ 
	\item $\LTLsquare \keep_e$
	\item ${\LTLsquare (\LTLcircle \decel_a \Rightarrow \decel_o)} \land {\LTLsquare(\LTLcircle \keep_a\Rightarrow \neg \accel_o)}$
\end{enumerate}
The objectives for \ego remain the same. \ahead is required not to accelerate in a curve, i.e., when \ahead notices a \bend two steps ahead, it may not accelerate two steps later, while in the curve. The car \other has to ensure not to get faster than \ahead. This is here expressed via a sufficient constraint on its acceleration.

We order processes $\ego>\other>\ahead$
and start the incremental synthesis with \ego. We have already seen elements of the set of \ego's assumption trees for the simpler setting of two cars in Sect.~\ref{sect:assumptions}. An assumption tree for the three car example will additionally be labeled with acceleration values for \ahead and the environmental input announcing bends $\bendannounce\in\{\bend, \nobend\}$. But since \ego depends on neither of them, they are of no relevance for the assumption tree, i.e., for each assumption tree in the two car setting there will be now an assumption tree where \ahead and \bendannounce are unconstrained. 

In the following we will illustrate the assumption propagation, starting with the exemplary assumption tree of ego depicted in Fig.~\ref{fig:sbs_strat+assm_keep}. 
Next, we consider a strategy annotated assumption tree of \other that matches with \ego's assumption tree of Fig.~\ref{fig:sbs_strat+assm_keep}..
For the sake of a simple illustration, we have chosen a strong assumption of \other about \ahead, that is annotated with a simple strategy for \other: The strategy annotated assumption tree, \saats(\other), given in Fig.~\ref{fig:assm_other}, describes the assumption that \ahead chooses $\accel_a$ at least every other time. 
The strategy $(\decel_o\accel_o)^{\omega}$ for \other is dominant strategy wrt. the assumption tree in Fig.~\ref{fig:assm_other}. 

Now, the \saats(\other) satisfies \ego's assumption \emph{"\other guarantees to choose $\neg \keep_o$ within its next two steps"}. 
A compatible dominant strategy for \ego is determined based on the propagated strategy annotated assumption tree. The match with \ego's strategy is highlighted in blue in Fig.~\ref{fig:assm_other}. The nodes are also annotated with \ego's acceleration choices. The numbers in brackets refer to the respective edge labels in the assumption tree in Fig.~\ref{fig:sbs_strat+assm_keep}. 
\begin{figure}
\centering
\includegraphics[scale=0.9]{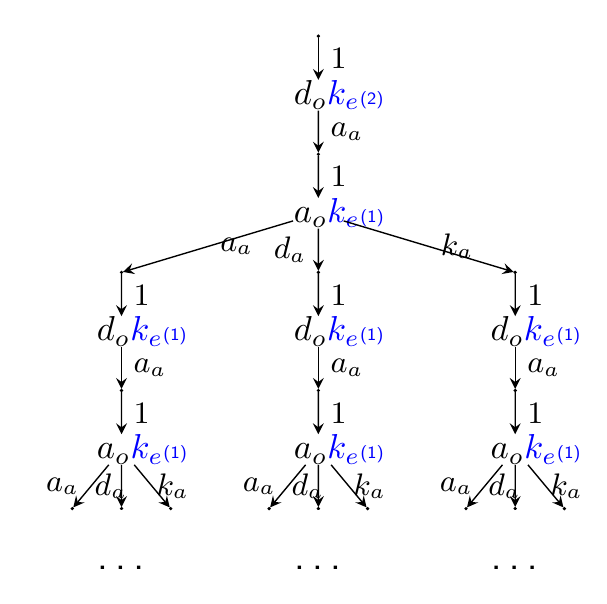}
\caption{A strategy annotated assumption tree of \other: \ahead is assumed to apply $\accel_a$ at least every other time. Other's dominant strategy is to alternate between $\decel_o$ and $\accel_o$. The tree satisfies \saats(\ego) of Fig.~\ref{fig:sbs_strat+assm_keep}.}\label{fig:assm_other}
\end{figure}

Intuitively, the weakest assumption for \other is \emph{"\ahead announces when it will do $\decel_a$ or $\keep_a$ next"} and a dominant strategy for \other is, to then choose the acceleration appropriately.

At this point the inputs \bendannounce and the acceleration of \ahead are open. Since the exemplary strategies do not refer to these, we again omitted these variables at the edge annotation in Fig.~\ref{fig:assm_other}.

Next and final element according to our process order is \ahead. 
So we choose a dominant strategy for \ahead, for instance a strategy that chooses $\decel_a$ in two steps time after the environment announces a curve two steps ahead, otherwise it chooses to accelerate. The strategies for \ego and \ahead are now determined based on the propagated (\ego,\other)-strategy annotated assumption trees, as illustrated in the previous step. Intuitively, when \ahead gets informed that a bend is two steps ahead, \other knows from the propagated strategy assumption tree that \ahead will decelerate in two steps time and chooses to decelerate at its the next step.
\end{exampleblock}
\section{Conclusion}\label{sect:conclusion}
Relying on cyber-physical systems increases our vulnerability towards technical failures caused by unpredicted emergent behavior. Understanding what pieces of information must be exchanged with whom, understanding what level of cooperation is necessary for the overall objectives --topics addressed in this paper-- constitute initial steps towards a rigorous design discipline for such systems. They also expose the vulnerability of the system towards intruders: the analysis clearly can be used to highlight the chaos, which can be inserted into the system by pretending to pass information and then not acting accordingly. The paper thus both provides an initial step towards cooperatively establishing systems objectives, as well as exposing its vulnerabilities. We will explore both avenues in our future research in the application domain of cooperative driver assistance systems.
\bibliographystyle{eptcs}
\bibliography{paper2}
\end{document}